\documentclass[preprint,aps,tightenlines,superscriptaddress]{revtex4}
\usepackage{graphicx}
\usepackage{bm}
\usepackage{epsfig}
\newif\ifpdf
\ifx\pdfoutput\undefined
\pdffalse 
\else
\pdfoutput=1 
\pdftrue
\fi
\newcommand{\bea}{\begin{eqnarray}}
\newcommand{\eea}{\end{eqnarray}}
\newcommand{\beq}{\begin{equation}}
\newcommand{\eeq}{\end{equation}}
\newcommand{\bay}{\begin{array}}
\newcommand{\eay}{\end{array}}
\begin{document}
\ifpdf
\DeclareGraphicsExtensions{.pdf, .jpg}
\else
\DeclareGraphicsExtensions{.eps, .jpg}
\fi
\preprint{ \vbox{\hbox{DUKE-TH-03-235} \hbox{hep-ph/0301187}  }}

\title{\phantom{x}\vspace{0.5cm} 
$1/N_c$ corrections to the $SU(4)_c$ symmetry for $L=1$ baryons:\\ The Three Towers
\vspace{0.5cm} }

\author{Dan Pirjol}
\affiliation{Department of Physics and Astronomy, Johns Hopkins University,
	Baltimore, MD 21218\footnote{dpirjol@pha.jhu.edu}}

\author{Carlos Schat}
\affiliation{Department of Physics, Duke University, Durham, 
NC  27708\footnote{schat@phy.duke.edu}}

\date{\today\\ \vspace{1cm} }

\begin{abstract}
In the large $N_c$ limit, the mass spectrum of the $L=1$ orbitally excited
baryons $N^*$ has a very simple structure, with states degenerate in pairs of spins 
$J=(\frac{1}{2},\frac32), (\frac{3}{2},\frac{5}{2})$, corresponding to irreducible 
representations (towers) of the contracted $SU(4)_c$ symmetry group.  
The mixing angles are completely determined in this limit.
Using a mass operator approach, we study $1/N_c$ corrections to this picture,
pointing out a four-fold ambiguity in the correspondence of the observed baryons 
with the large $N_c$ states. For each of the four possible assignments, we
fit the coefficients of the quark operators contributing to the
mass spectrum to $O(N_c^{-1})$. We comment on the implications of our results for the
constituent quark model description of these states.
\end{abstract}

\maketitle
\newpage

\section{Introduction}

Ever since their discovery, the negative-parity excited baryons have
been a testing ground for the quark model, as their complex 
mass spectrum makes it possible to distinguish among different models of 
quark-quark interactions. In the quark model, the $L=1$ negative parity baryons 
fall into the ${\bf 70}$ representation of $SU(6)$, which contains the $({\bf 1}, 2)$,
$({\bf 10}, 2)$, $({\bf 8}, 2)$ and $({\bf 8}, 4)$ of $SU(3)_{\rm fl}\otimes SU(2)_{\rm sp}$
\cite{su6,mix}.
The finer details of the spectrum were first studied in a constituent quark model 
by Isgur and Karl \cite{IsgKa,CaIsg}. Recently, the flavor dependence of the quark spin 
interactions has been a matter of some debate, related to their dynamical origin
as arising from constituent gluons or pions exchange \cite{OPE,HG,Riska}. 

In this paper we focus on a different, model-independent description of these states,
which was proposed only relatively recently 
\cite{DJM1,DJM2,CGKM,JG,PY1,PY2,CCGL1,CCGL,CC,SGS,su3prd}. 
This is based on the large $N_c$ limit \cite{1,2},
and makes use of the contracted $SU(2N_f)$ symmetry of QCD in this limit \cite{DJM2}.
There are two equivalent ways of extracting the predictions of this symmetry.
The first method is algebraic, and makes use of commutation relations of operators 
representing physical quantities such as axial currents and hadron masses, which are
constrained by so-called consistency conditions.
Although somewhat cumbersome and
difficult to extend beyond the leading order in $1/N_c$, such an approach has 
been applied to the study of the mass spectrum and strong decays of the
excited light baryons in \cite{PY1,PY2}. These states were predicted to
fall into irreducible representations of the contracted $SU(2N_f)$ algebra,
with nontrivial implications for the mass spectrum and mixing angles of the
excited baryons.

A somewhat different approach to large $N_c$ baryons is based on the
quark operators method. In this approach the representations of the contracted
$SU(2N_f)$ algebra are constructed in terms of a `quark' basis 
$|\frac12, s_3 \rangle\otimes| \frac12, i_3 \rangle$. Although similar to the
one-quark basis states used in the constituent quark model (CQM), we stress 
that this is a purely mathematical device and does not imply any of the dynamical
assumptions of the CQM. 
Any physical quantity is further written as an expansion in $n$-body operators acting 
on the `quark' states, ordered according to $n$. This can be organized as
an expansion in $1/N_c$, with a finite number of independent operators
contributing at any given order in $1/N_c$. This method for extracting the
predictions of the large $N_c$ limit has the advantage of being more systematic 
and easier to extend to subleading orders in $1/N_c$.

The mass spectrum of the excited baryons has been studied using this approach in a
series of papers \cite{CCGL1,CCGL,SGS,su3prd}, including operators of order up to 
and including $1/N_c^2$.
In these papers the coefficients of these operators have been extracted from a 
fit to the observed masses, yielding a very specific pattern of the dominant operators.

In this paper we reanalyze the $1/N_c$ expansion for the masses of the nonstrange 
excited baryons. 
Guided by the expected structure of the large $N_c$ mass spectrum, we
present fits for the coefficients of the mass operator, order by order in
$N_c$. We point out a four-fold ambiguity in the correspondence of the physical
states with the large $N_c$ states, and the special status of the excited $\Delta^*$
states in the $N_c$ expansion.  At leading order in $N_c$ only three
operators contribute to the mass operator, and we discuss the four equivalent 
fits to their coefficients, one of which turns out to be ruled out.
Including $1/N_c$ corrections allows eight operators to contribute to the mass 
matrix. Considering only data on $N^*$ masses, the coefficients of these eight
operators can be determined as functions of the two mixing angles, up to an
additional continuous ambiguity. We present  restrictive constraints
on the allowed set of coefficients, first from dimensional analysis and then by 
including data on the  excited 
$\Delta^*$ states.  We compare the results of our fits to those
previously discussed in the literature.

\section{Formalism}

The spin-flavor structure of the orbitally excited $L=1$ baryons for arbitrary 
$N_c$ can be conveniently described in terms of quark quantum numbers. The
quark basis states can be labeled as $|J,I,S; m_J, I_3\rangle$ where $S$ is the 
quarks' spin,
$I$ their isospin and $\vec J = \vec S+\vec L$ the total baryon spin. The spin-isospin
wave function of a nonstrange $L=1$ orbitally excited baryon has mixed spin-flavor 
symmetry. This implies that for a given isospin
$I$, the spin $S$ takes all values compatible with $|S-I| \leq 1$ except 
for $S=I=N_c/2$. The most general state $|JIS\rangle$ can be constructed by 
adding one excited quark to a symmetric 'core' of quantum numbers $|S_c = I_c\rangle$, 
such that  $\vec S = \vec S_c + \vec s$, $\vec I = \vec I_c + \vec i$
with $s = i = \frac12$.

In the large $N_c$ limit the mass eigenstates obtained from this construction
fall into irreducible representations of the contracted $SU(2N_f)$ algebra \cite{DJM2}.
These are infinite towers of states, labelled by an integer or half-integer
$T$, and contain all possible states satisfying the condition $|J-I|\leq T$.
All the states belonging to a tower are degenerate in mass, and the separation
between towers is of order $O(N_c^0)$ in the large $N_c$ limit. On the other hand,
mass splittings within towers  are small, and
appear first at order $O(1/N_c)$. The tower states 
$|JIT\rangle$ are related to the quark model states $|JIS\rangle$ by a simple 
recoupling relation \cite{PY1} \footnote{The states $|JIS\rangle$ appearing on 
the r.h.s
are related to those of \cite{CCGL} as $|\frac12 \frac12 \frac12 \rangle = - N_{1/2},
|\frac12 \frac12 \frac32 \rangle = N'_{1/2}, |\frac32 \frac12 \frac12 \rangle = - N_{3/2},
|\frac32 \frac12 \frac32 \rangle = N'_{3/2}$.}
\bea\label{mixing}
|JIT\rangle = (-)^{I+1+L+J}\sum_S \sqrt{(2S+1)(2T+1)}
\left\{
\begin{array}{ccc}
I & 1 & S \\
L & J & T \\
\end{array}\right\}
|JIS\rangle\,.
\eea
This completely determines  the mixing matrix of the $L=1$ excited baryons in the
large $N_c$ limit, in the absence of configuration mixing \footnote{We thank Rich
Lebed for discussions on this point. See also \cite{CoLe}.}.

The resulting mass spectrum for the nonstrange $P-$wave baryons is shown in 
Fig.~1(b).
There are five $I=1/2$ states ($N-$type) which correspond to quark model
states of spin $S=1/2, 3/2$, and eight $I=3/2$ states ($\Delta-$type, not shown), 
corresponding to quark spin $S=1/2,3/2$ and $5/2$. This covers all the observed $N$ 
type states, but due to the finite value of $N_c=3$ in the real world, only two 
$I=3/2$ states are present. This will be seen to limit the predictive power of the 
large $N_c$ approach for the $\Delta-$type excited baryons.

\begin{figure}[t!]
  \includegraphics[width=4in]{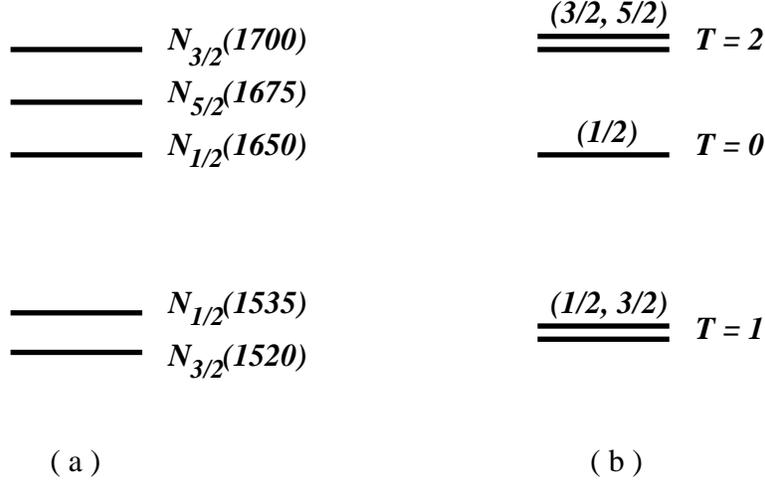} 
{\caption{The mass spectrum of the nonstrange $L=1$ orbitally excited
nucleons. (a) the observed
states  and (b) in the large $N_c$ limit  these states
fall into three towers labelled by the quantum number $T=0,1,2$.}}
\end{figure}

In the following we show how the large $N_c$ predictions are reproduced
in the quark operators language. This will allow us to include also symmetry
breaking to this picture, appearing at order $1/N_c$.
The mass matrix of the $L=1$ baryons can be written as a sum of operators
acting on the quark basis as
\bea\label{massop}
\hat M = \sum_{k=0}^{N_c} \frac{1}{N_c^{k-1}} C_k {\cal O}_k
\eea
with ${\cal O}_k$ a $k$-body operator. Both the coefficients $C_k$ 
and the matrix elements of the operators on baryon states $\langle {\cal O}_k\rangle$
have power expansions in $1/N_c$ with coefficients determined by nonperturbative
dynamics
\bea
C_k = \sum_{n=0}^\infty \frac{1}{N_c^n} C_k^{(n)}\,, \qquad
\langle {\cal O}_k \rangle = \sum_{n=0}\frac{1}{N_c^n}\langle {\cal O}_k 
\rangle^{(n)} \,.
\eea
The natural size for the coefficients $C_k^{(n)}$ is $\Lambda/N_c^n$, with 
$\Lambda \sim 500 $ MeV.
The basic building blocks for constructing the operators ${\cal O}_k$
are $s^i, \tau^a, g^{ia}$ (acting on the excited quark), 
$S_c^i, G_c^{ia}$ (acting on the core),
and operators acting on the orbital degrees of freedom 
$l^i, l^{(2)ij}=\frac12 \{l^i, l^j\} - \frac13 l^2 \delta^{ij}$.

We will use the operator basis introduced in \cite{CCGL}, which contains
the following operators, ordered according to their leading power in $1/N_c$.
At leading order in $N_c$ only three operators contribute to the mass matrix,
given by
\bea
{\cal O}_1 = N_c {\bf 1}\,,\qquad
{\cal O}_2 = l^i s^i\,,\qquad
{\cal O}_3 = \frac{3}{N_c}l^{(2)ij} g^{ia} G_c^{ja}\,.
\eea
One small technical difference with the operators in \cite{CCGL} is that we
define ${\cal O}_3$ with an added factor of 3. With this redefinition, its
matrix elements are not anomalously small, which is required for a consistent
dimensional analysis for its coefficients \cite{SGS,su3prd}.
At subleading order   $O(N_c^{-1})$ five additional operators
start contributing, which can be chosen as
\bea
& &{\cal O}_4 = l s + \frac{4}{N_c+1} l t G_c\,,\qquad
{\cal O}_5 = \frac{1}{N_c} l S_c\,,\qquad 
{\cal O}_6 = \frac{1}{N_c} S_c S_c\,,\\
& &\qquad\qquad {\cal O}_7 = \frac{1}{N_c} s S_c\,,\qquad
{\cal O}_8 = \frac{1}{N_c} l^{(2)} s S_c\,.\nonumber
\eea
These operators have a direct physical interpretation in the quark model
in terms of one- and two-body quark-quark  couplings. 

We start by examining the mass spectrum of the $N^*$ states in the 
large $N_c$ limit. Keeping only the ${\cal O}_{1,2,3}$ operators contributing 
at $O(N_c^0)$, one finds by direct diagonalization of the mass matrix
the mass eigenstates in the large $N_c$ limit as linear combinations of
the quark model $N_{1/2}, N'_{1/2}$ states (apart from the redefinition of
${\cal O}_3$, we use everywhere the notations of \cite{CCGL})
\bea\label{Delta0}
|T = 0\,, J=\frac12 \rangle &=& 
\frac{1}{\sqrt3} N_{1/2} + \sqrt{\frac23} N'_{1/2}\\
\label{Delta1}
|T = 1\,, J=\frac12 \rangle &=& 
-\sqrt{\frac23} N_{1/2} + \frac{1}{\sqrt3} N'_{1/2}
\eea
with masses
\bea\label{M0}
M_0^{(0)} &=& N_c C_1^{(0)} - C_2^{(0)} - \frac{5}{8} C_3^{(0)}\\
\label{M1}
M_1^{(0)} &=& N_c C_1^{(0)} - \frac12 C_2^{(0)} + \frac{5}{16} C_3^{(0)}\,.
\eea

A similar diagonalization of the mass matrix for the $J=\frac32$ $N^*$ states gives the
eigenstates
\bea\label{Delta1'}
|T = 1\,, J=\frac32 \rangle &=& 
\frac{1}{\sqrt6} N_{3/2} + \sqrt{\frac56} N'_{3/2}\\
\label{Delta2}
|T = 2\,, J=\frac32 \rangle &=& 
-\sqrt{\frac56} N_{3/2} + \frac{1}{\sqrt6} N'_{3/2}
\eea
with masses
\bea
M_1^{(0)} &=& N_c C_1^{(0)} - \frac12 C_2^{(0)} + \frac{5}{16} C_3^{(0)}\\
\label{M2}
M_2^{(0)} &=& N_c C_1^{(0)} + \frac12 C_2^{(0)} - \frac{1}{16} C_3^{(0)}\,.
\eea
The $N_{5/2}$ state does not mix and has the mass $M_2^{(0)}$.

These results make the tower structure shown in Fig.~1(b) explicit, with four
of the states degenerate in pairs $(I,J)=(\frac12,\frac12), (\frac12,\frac32)$ 
(corresponding to the
$T=1$ tower) and $(I,J)=(\frac12,\frac32), (\frac12, \frac52)$ (corresponding 
to the $T=2$
tower), respectively. Also, the mixing matrices agree with the expected recoupling
relation (\ref{mixing}) and do not depend on the explicit values of the coefficients
$C_{2,3}^{(0)}$, the reason being that $ {\cal O}_2$ , ${\cal O}_3$ commute in the large $N_c$ limit.
The mass splittings between the towers are determined  by the
coefficients $C_2^{(0)}$ and $C_3^{(0)}$.

There is a discrete ambiguity in the correspondence of the large $N_c$ tower 
states and the 
observed $N^*$ excited nucleons. Taking the number of colors $N_c$ to be infinitely 
large, the states  arrange themselves into towers
as described above. The observed mass values (see Fig.~1)  suggest  assigning
the two lowest lying states $N(1520)$ and $N(1535)$ into the $T=1$
tower, and then group the states of spins 3/2 and 5/2 into the $T=2$\
tower. The remaining $J=1/2$ state $N(1650)$ would belong to the $T=0$ tower.
This was the assignment suggested in \cite{PY2} as it comes closest to the
large $N_c$ limit picture.

\bea\nonumber
\begin{array}{c|c|c|c}
\hline
\mbox{Assignment} & T = 0 & T = 1 & T = 2 \\
\hline
\hline
\mbox{\# 1} & N_{1/2}(1650) & \{N_{1/2}(1535), N_{3/2}(1520)\} &  
\{N_{3/2}(1700), N_{5/2}(1675)\} \\
\mbox{\# 2} & N_{1/2}(1535) & \{N_{1/2}(1650), N_{3/2}(1520)\} &  
\{N_{3/2}(1700), N_{5/2}(1675)\} \\
\mbox{\# 3} & N_{1/2}(1535) & \{N_{1/2}(1650), N_{3/2}(1700)\} &  
\{N_{3/2}(1520), N_{5/2}(1675)\} \\
\mbox{\# 4} & N_{1/2}(1650) & \{N_{1/2}(1535), N_{3/2}(1700)\} &  
\{N_{3/2}(1520), N_{5/2}(1675)\} \\
\hline
\end{array}
\eea
\begin{quote}
Table 1. The four possible assignments of the observed nonstrange excited
baryons into large $N_c$ towers with $T = 0, 1, 2$.
\end{quote}

However, this is not the most general possible assignment. Allowing for potentially
large $1/N_c$ corrections, there are four possible assignments of the observed
states into towers, defined in Table 1.

Each of these assignments leads to a different picture in the
large $N_c$ limit, and to different predictions for the properties of these
states. For example, the large $N_c$ predictions for ratios of strong decay 
amplitudes depend on the tower assignment of the states \cite{PY1}.
In the next Section we determine the coefficients $C_{1-3}^{(0)}$ of the 
operators appearing in (\ref{massop}) by imposing the constraint that in the
large $N_c$ limit the physical states go over into the towers corresponding to
each assignment.

The status of the excited $\Delta^*$ states appears to be special.
In the large $N_c$ limit, there are eight $I=3/2$ ($\Delta-$type) states, which
arise in the quark model from states with quark spin $S=1/2,3/2$ and $5/2$.
Coupling the spin $S$ with the orbital momentum $L=1$ gives 2 $\Delta^*_{1/2}$,
3 $\Delta^*_{3/2}$, 2 $\Delta^*_{5/2}$ and one $\Delta^*_{7/2}$ states, where the subscript
denotes the total $\vec J=\vec S+\vec L$ hadron spin. Diagonalizing the mass 
matrices of these
states, one expects the mass eigenstates to arrange themselves again into towers,
degenerate with the $N^*$ states, as follows: one $\Delta^*_{3/2}$ state with mass
$M_0^{(0)}$ corresponding to the $T=0$ tower, 3 states $\Delta^*_{1/2,3/2,5/2}$
degenerate with mass $M_1^{(0)}$ (the $T=1$ tower), and 4 states
$\Delta^*_{1/2,3/2,5/2,7/2}$ degenerate with mass $M_2^{(0)}$ (the $T=2$ tower).

The real world is very different: at $N_c=3$ only two $\Delta^*$ states exist, with 
spins $J=1/2, 3/2$. 
Consider for example the mass matrix of the two $\Delta^*_{1/2}$ states for arbitrary 
$N_c \geq 5$.
Expressed in the basis of the quark model states 
$|JIS\rangle = |\frac12\frac32 \frac12\rangle$
and $|\frac12\frac32 \frac32\rangle$ it has the form
\bea
{\cal M}_{\Delta^*_{1/2}} = \left(
\begin{array}{cc}
M_1(N_c) & m_{12}(N_c) \\
m_{12}(N_c) & M_2(N_c) \\
\end{array}
\right)\,.
\eea
The matrix elements $M_{1,2}(N_c), m_{12}(N_c)$ have power expansions in $1/N_c$ 
\cite{CCGL}
\bea\label{Delta12}
M_1(N_c) &=& N_c C_1(N_c) + \frac13 C_2(N_c) +O(N_c^{-1})\,,\\
M_2(N_c) &=& N_c C_1(N_c) + \frac{15-2N_c}{6N_c} C_2(N_c)+
\frac{2N_c^2+2N_c-15}{8N_c^2}C_3(N_c) + O(N_c^{-1})\,,\\
m_{12}(N_c) &=& \frac{1}{6}\sqrt{\frac{5(N_c-3)}{N_c}} \left( C_2(N_c) - 
\frac{3(2N_c + 5)}{16 N_c} C_3(N_c)\right) + O(N_c^{-1})\,.
\eea
Denoting the eigenvalues of this matrix with $\lambda_{1,2}(N_c)$,
they can be also
written as expansions in $1/N_c$, $\lambda_{1,2}(N_c) = N_c C_1(N_c) + O(N_c^0) + \cdots$.
It is easy to check that the eigenvalues in the large $N_c$ limit are identical
with the tower masses $M_{1,2}^{(0)}$ given in (\ref{M1}), (\ref{M2}).
At $N_c=3$ the `phantom' state  $S=3/2$ disappears, and 
its mixing with the `physical' $S=1/2$ state vanishes. 
The mass of the observed state $\Delta^*_{1/2}$ is identified with the diagonal matrix 
element $M_1(N_c=3)$. 

The meaning of the large $N_c$ expansion for this case is less clear than for the
$N^*$ states, since the $N_c=3$ and large $N_c$ sets of states are completely different.
In particular, it is not completely clear that
the large $N_c$ expansion can predict the masses of the $\Delta^*$ states with 
corrections parametrically suppressed by $1/N_c$, as for the $N^*$ states.
For example, the $O(N_c^0)$ term in the expansion of the $\Delta^*_{1/2}$ mass
(\ref{Delta12}) differs from both tower masses (\ref{M1}), (\ref{M2})
(to which it goes as $N_c\to\infty$) by terms of $O(N_c^0)$.

For these reasons we will perform our analysis below in two steps.
First, we exclude the $\Delta^*$ states and explore the form of the
most general constraints which can be obtained from the $N^*$ alone.
Then we add them in a second step, which allows us to compare our results
with previous fits in the literature, where they are taken into account.

\section{Numerical results and fits}

The general form of the mass matrix with $N_c=3$ is rather complicated and involves
mixing among states with the same quantum numbers. Therefore a direct analysis is 
not very transparent, and we present first a simplified fit at leading order in
$1/N_c$. For this purpose, we keep only the $O(N_c,N_c^0)$ terms in the mass matrix, 
which come from the unit operator ${\cal O}_1$ and the leading terms
in the expansion of the matrix elements of ${\cal O}_{2,3}$.
In a second step we allow for the contributions of all $1/N_c$ operators, which 
can reproduce the finer structure within the towers.

We start by determining the values of the coefficients $C_{1,2,3}^{(0)}$ in the 
large $N_c$ limit. 
The large $N_c$ mass eigenvalues and eigenstates have been presented in 
Eqs.~(\ref{M0}), (\ref{M1}) and (\ref{M2}).  For each assignment, we fitted the 
coefficients $C_{1,2,3}^{(0)}$ to the observed $N^*$ masses. 
We use the experimental values for the masses given in Table V of \cite{CCGL}.
This gives the results for $C_{1-3}^{(0)}$ shown in Table 2 for each
of the four possible assignments, together with the mixing angles 
$\theta_{N1}, \theta_{N3}$, which are fixed to the large $N_c$ values
as given in (\ref{Delta0}), (\ref{Delta1}), (\ref{Delta1'}), (\ref{Delta2}). 
We define the mixing angles as in \cite{CGKM}
\bea
\left(
\begin{array}{c}
N(1535) \\
N(1650) \\
\end{array}
\right) = \left(
\begin{array}{cc}
\cos \theta_{N1} & \sin \theta_{N1} \\
-\sin \theta_{N1} & \cos \theta_{N1} \\
\end{array}
\right)
\left(
\begin{array}{c}
N_{1/2} \\
N'_{1/2} \\
\end{array}
\right)
\eea
and
\bea
\left(
\begin{array}{c}
N(1520) \\
N(1700) \\
\end{array}
\right) = \left(
\begin{array}{cc}
\cos \theta_{N3} & \sin \theta_{N3} \\
-\sin \theta_{N3} & \cos \theta_{N3} \\
\end{array}
\right)
\left(
\begin{array}{c}
N_{3/2} \\
N'_{3/2} \\
\end{array}
\right)\,.
\eea
As mentioned in Sec.~II, the best fit is obtained 
for the assignment \#1,
for which the observed $N^*$ mass spectrum comes closer to the tower structure 
in Fig.1(b).

These results must satisfy an additional constraint, following
from the no-crossing property of the eigenstates with the same quantum
numbers. Consider for example the masses of the $J=1/2$ states as functions of
$1/N_c$. They can not cross when $N_c$ is taken from 3 to infinity, which means
that the correspondence of the physical $N_{1/2}$ states with the large $N_c$
towers is fixed by the relative ordering of the $T=0,1$ towers. A similar 
argument can be given for the $J=3/2$ states which fall into the $T=1,2$ towers.

\bea\nonumber
\begin{array}{|c|rrr|c|ccc|cc|rr}
\hline
\mbox{Assignment} & N_c C_1^{(0)} & C_2^{(0)} & C_{3}^{(0)} &  
\chi^2/{\rm d.o.f.} & M_0 & M_1 & M_2  & \theta_{N1} & \theta_{N3}\\
\hline
\hline
\mbox{\# 1} & 1625 \pm 8 & 83 \pm 14  & -188 \pm 28 &  0.38  & 1660 & 1525 & 1679 & 2.53 & 1.15 \\
\mbox{\# 2} & 1617 \pm 8 & 115 \pm 14 &  -57 \pm 27 &  20.3  & 1538 & 1542 & 1679 & 0.96 & 1.15 \\
\mbox{\# 3} & 1615 \pm 9 & -12\pm 16  &  142 \pm 38 &  94.1  & 1538 & 1666 & 1601 & 0.96 & 2.72 \\
\mbox{\# 4} & 1592 \pm 9 &   2\pm 16  & -111 \pm 40 &  98.5  & 1660 & 1557 & 1601 & 2.53 & 2.72 \\
\hline
\end{array}
\eea
\begin{quote}
Table 2. The values of the coefficients $C_{1-3}^{(0)}$ (in MeV), the masses of the towers
$M_{0,1,2}$ (in MeV) and the mixing angles (in rad) following from the
large $N_c$ fit, for  each of the 4 assignments. 
\end{quote}

Taken together with the ordering of the observed hadron masses, this gives the
following correspondence of the four assignments with the ordering of the
tower masses: 
\bea
& &\#1: \qquad \{M_0\,, M_2\} > M_1,\\
& &\#2: \qquad M_2 > M_1 > M_0\,,\nonumber\\
& &\#3: \qquad M_1 > \{M_0\,, M_2\}\,,\nonumber \\
& &\#4: \qquad M_0 > M_1 > M_2\,.\nonumber
\eea 
Furthermore, since the relative ordering of the tower masses is given by 
$C_{2,3}^{(0)}$, there is a 
unique assignment corresponding to each set of these coefficients. This correspondence
is shown in a graphical form in Fig.~2, together with the results for the coefficients
$C_{2,3}^{(0)}$ following from the fits to the physical masses in Table 2. 

The plot in Fig.~2 can be used to further constrain the values of the coefficients
$C_{2,3}^{(0)}$.
Note that the region for these coefficients
obtained for assignment \#2 crosses partially into the region corresponding to \#1.
This restricts further the range of values for $C_{2,3}^{(0)}$ corresponding to
\#2. Also, the fit results corresponding to 
\#4 has no overlap with the area corresponding to this assignment, which means that 
the assignment \#4 is ruled out at leading order in $N_c$.

\begin{figure}[t!]
  \includegraphics[width=3in]{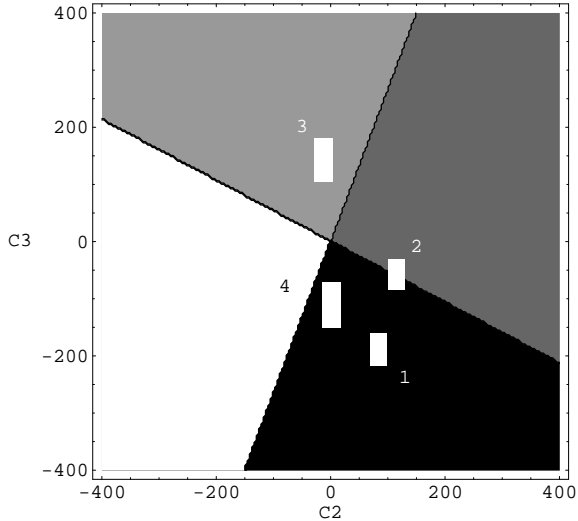}
{\caption{
Regions in the $(C_2^{(0)}, C_3^{(0)})$ plane corresponding to 
each assignment. Each area corresponds to one assignment: \#1 (black), \#2 (dark
grey), \#3 (light grey), \#4 (white). The line separating the assignments
\#3,\#4 and \#2,\#1 is given by $3C_3^{(0)}+\frac85 C_2^{(0)}=0$, and the line separating
the regions for assignments \#2,\#3 and \#1,\#4 is $3C_3^{(0)} - 8 C_2^{(0)} = 0$.
The strips show the results for the
coefficients $C_{2,3}^{(0)}$ following from the large $N_c$ fit in Table 2, 
for each assignment.}}
\end{figure}

The results of this analysis can be compared directly with a 3-parameter fit
performed in \cite{CCGL}, where only the operators ${\cal O}_{1,2,3}$ were included
(see Table V in \cite{CCGL}). The only differences are the inclusion of the
$\Delta^*$ states in \cite{CCGL} and the fact that the authors of 
\cite{CCGL} use in this fit
the matrix elements of the operators ${\cal O}_{1,2,3}$ at $N_c=3$, while we evaluate
them in the large $N_c$ limit. The coefficients obtained in \cite{CCGL} are 
$N_c C_1 = 1626\pm 6$, $C_2 = 75\pm 9$, $C_3 = -146\pm 17$ (in MeV), and come closest to 
our assignment \#1. However, when all operators up to order $O(1/N_c)$ are included, 
the coefficients obtained in \cite{CCGL} (reproduced here in Table 3) come closest
to our assignment \#3. 

\pagebreak

\bea\nonumber
\begin{array}{ccc|ccccc}
\hline
 N_c C_1 & C_2 & C_3 & C_4 & C_5 & C_6 & C_7 & C_8 \\
\hline
\hline
 1389 \pm 60 & -36\pm 41 & 123\pm 69 & 87\pm 97 & 86\pm 80 &
438 \pm 102 & -40\pm 74 & 48\pm 172 \\
\hline
\end{array}
\eea
\begin{quote}
Table 3. Results for the coefficients $C_k$ (in MeV) obtained from the fit of
\cite{CCGL}, using as input parameters the
excited $N$, $\Delta^*$ and mixing angles $(\theta_{N1},\theta_{N3}) = (0.61,3.04)$ (in rad).
Note that our $C_3$ is related to that used in \cite{CCGL} by $C_3 = C_3^{\rm CCGL}/3$.
\end{quote}

Beyond leading order in $1/N_c$ there are three types of symmetry breaking corrections 
which introduce deviations from the tower picture:
\begin{enumerate}
\item Subleading corrections $C_{1-3}^{(1)}$ in the coefficients of the operators 
${\cal O}_{1-3}$.

\item $O(1/N_c)$ subleading corrections to the matrix 
elements of the operators ${\cal O}_{2-3}$.

\item Leading contributions from the matrix elements of the ${\cal O}_{4-8}$ operators.

\end{enumerate}
While type 1. corrections just shift the relative position of the towers,
the remaining two types of corrections introduce splittings among the tower states.
We will include subleading corrections to the matrix elements of the
operators ${\cal O}_{1-8}$ by evaluating them at the physical value $N_c=3$.

Using as inputs the five masses of the $N^*$ states, it would appear at the first 
sight that there
is a 3-dimensional manifold of solutions for the coefficients $C_{1-8}$ of the
full set of $O(1/N_c)$ operators. In fact this is reduced to a 2-dimensional set of
solutions because of an accidental relation among the matrix elements of the
mass operators with $N_c=3$. Namely, when restricted to the subspace of the 
$N^*$ states, a certain linear combination of ${\cal O}_{6,7}$ turns out to be 
equivalent to the unit operator
\bea\label{accident}
\langle {\cal O}_6\rangle_{N_c=3} - \langle {\cal O}_7\rangle_{N_c=3} =
\frac12 \cdot {\bf 1}\,.
\eea
This relation implies an ambiguity in the coefficients $C_{1,6,7}$ beyond leading
order in $1/N_c$, which can be summarized by noting that 
the mass spectrum of the $N^*$ states is invariant under the simultaneous changes
\bea\label{ambig}
N_c C_1 \to N_c C_1 + \delta\,,\quad C_6 \to C_6 - 2\delta\,,\quad 
C_7\to C_7 + 2\delta\,.
\eea
In other words, only the two combinations $N_c C_1^{\rm eff} = N_c C_1 - \frac12 C_7$ 
and $C_6^{\rm eff} = C_6 + C_7$ can be determined unambiguously from the $N^*$ masses. 
Of course, the relation (\ref{accident}) is not true anymore when considering also the 
$\Delta^*$ states, which can be used to resolve the ambiguity (\ref{ambig}).
However, as discussed in Sec.~II, due to the incomplete structure of the towers
in the $I=3/2$ sector at $N_c=3$, the mass spectrum 
of the observed $\Delta^*$ states does not have the correct large $N_c$ limiting behaviour. 
Therefore, we will keep our discussion as general as possible, and present separate
results with and without including the $\Delta^*$s.

To present the results of our numerical analysis for the coefficients $C_{1-8}$,
we need to pick a parameterization of the 2-dimensional manifold of 
solutions. A particularly convenient choice of the coordinates on this manifold
are the mixing angles $(\theta_{N1}, \theta_{N3})$, which will be left
completely arbitrary. Using as inputs the masses of the five $N^*$ states
one finds, at each point in the $(\theta_{N1}, \theta_{N3})$ plane, one set of
coefficients $C_{1,6}^{\rm eff}, C_{2-5,8}$. The analytical expressions for these
coefficients can be found in the Appendix.

\begin{figure}[t!]
  \includegraphics[width=3in]{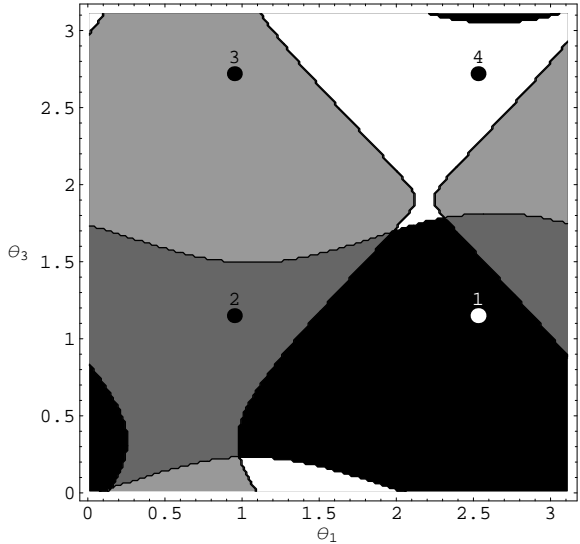} 
{\caption{The different regions in the $(\theta_{N1}, \theta_{N3})$ plane (in rad) corresponding to
the four assignments in Table I, determined as explained in the text. The black region 
- \# 1, dark grey - \# 2, light grey - \# 3, white - \# 4. The four dots shown 
correspond to the values of the mixing angles in the large $N_c$ limit.}}
\end{figure}

It is natural to ask if there is a unique association of a particular solution
at a given point $(\theta_{N1}, \theta_{N3})$ with one of the four assignments 
in Table 1. As noted above, the assignment which is realized depends on the
large $N_c$ values of the cofficients $C_{2,3}^{(0)}$ - see Fig.~2. Requiring
that the extracted values of $C_{2,3}(\theta_{N1}, \theta_{N3})$ lie inside the
regions in Fig.~2 associated with each assignment, one finds the
corresponding partition of the entire $(\theta_{N1}, \theta_{N3})$ plane into regions
shown in Fig.~3. This procedure assumes that the $1/N_c$ corrections to the
coefficients $C_{2,3}^{(0)}$ are negligible; including them will change slightly
the boundaries of the regions in Fig.~3.

Without additional input, the individual coefficients $C_{1-8}$ can not be
determined from this $1/N_c$ analysis, apart from the rather loose absolute 
bounds (obtained by varying $\theta_{N1,N3}$ over their entire range $[0,\pi)$): 
$1374 < N_c C^{\rm eff}_1  < 1770$,
$-148 < C_2 < 164$, 
$-326 < C_3 < 363$, 
$-169 < C_4 < 223$, 
$-169 < C_5 < 342$, 
$-251 < C_6^{\rm eff} < 463$,
$-638 < C_8 < 775$ (in MeV).

A more precise determination becomes possible if we impose constraints on the 
parameter space from dimensional analysis.
As discussed in Sec.~II, the natural size of the coefficients is $C_k^{(n)} 
\sim \Lambda/N_c^n$.
In particular, this means that the coefficients $C_{2,3}$ can
not be too different from their large $N_c$ values extracted in the previous step
(see Table 2). By simple power counting, one expects the $1/N_c$ corrections to these
coefficients to be of order $\Lambda/N_c \sim 150$ MeV.
Restricting the coefficients $C_{1-3}$ singles out a solution in the 
$(\theta_{N1},\theta_{N3})$ plane, resolving the ambiguity (\ref{ambig}) and fixing the
remaining coefficients $C_{4-8}$.

\begin{figure}[t!]
  \includegraphics[width=2in]{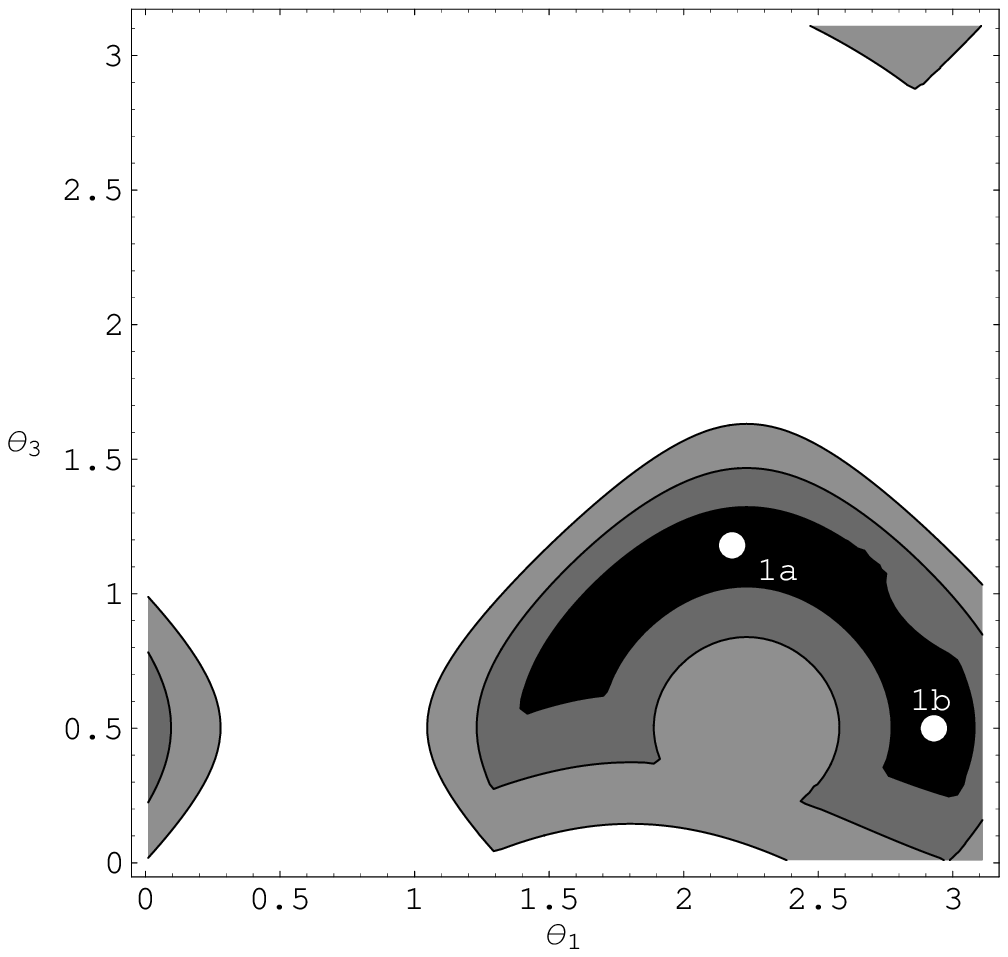}\hspace{1cm}
  \includegraphics[width=2in]{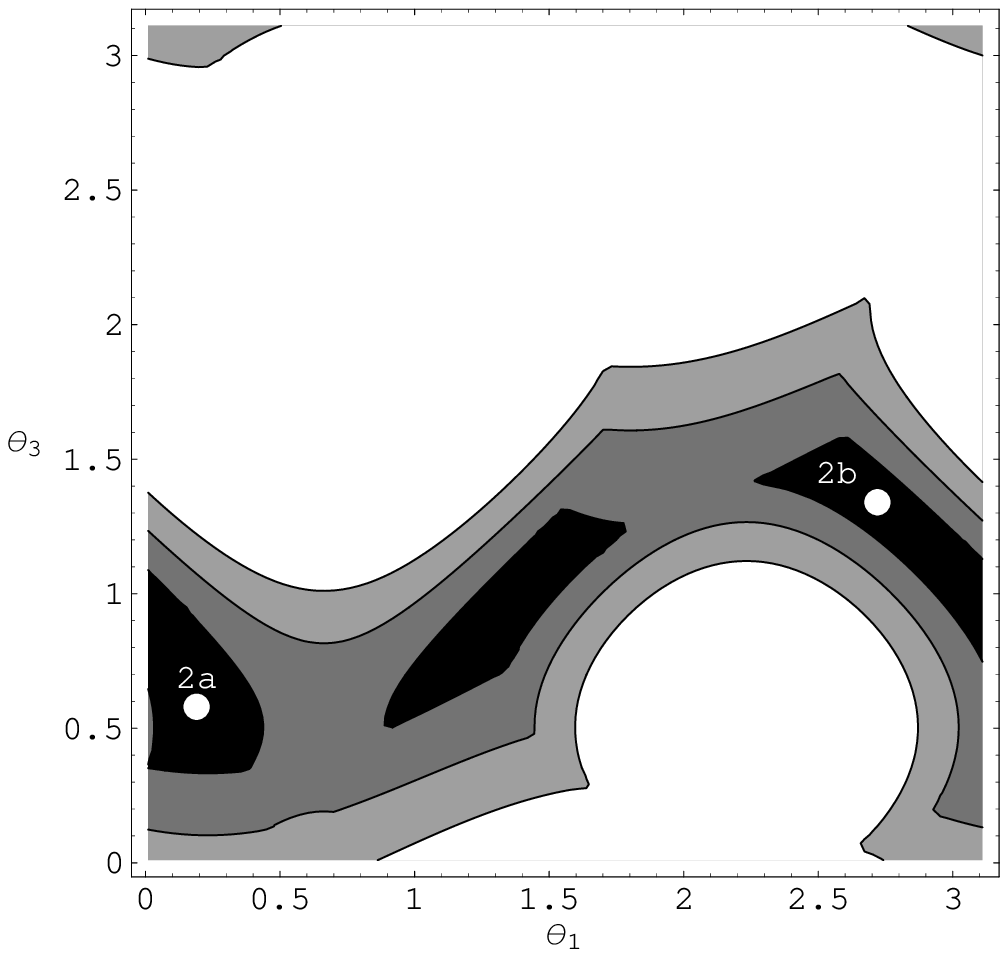}\\
\hspace{1cm} (a) \hspace{5cm} (b) \\
  \includegraphics[width=2in]{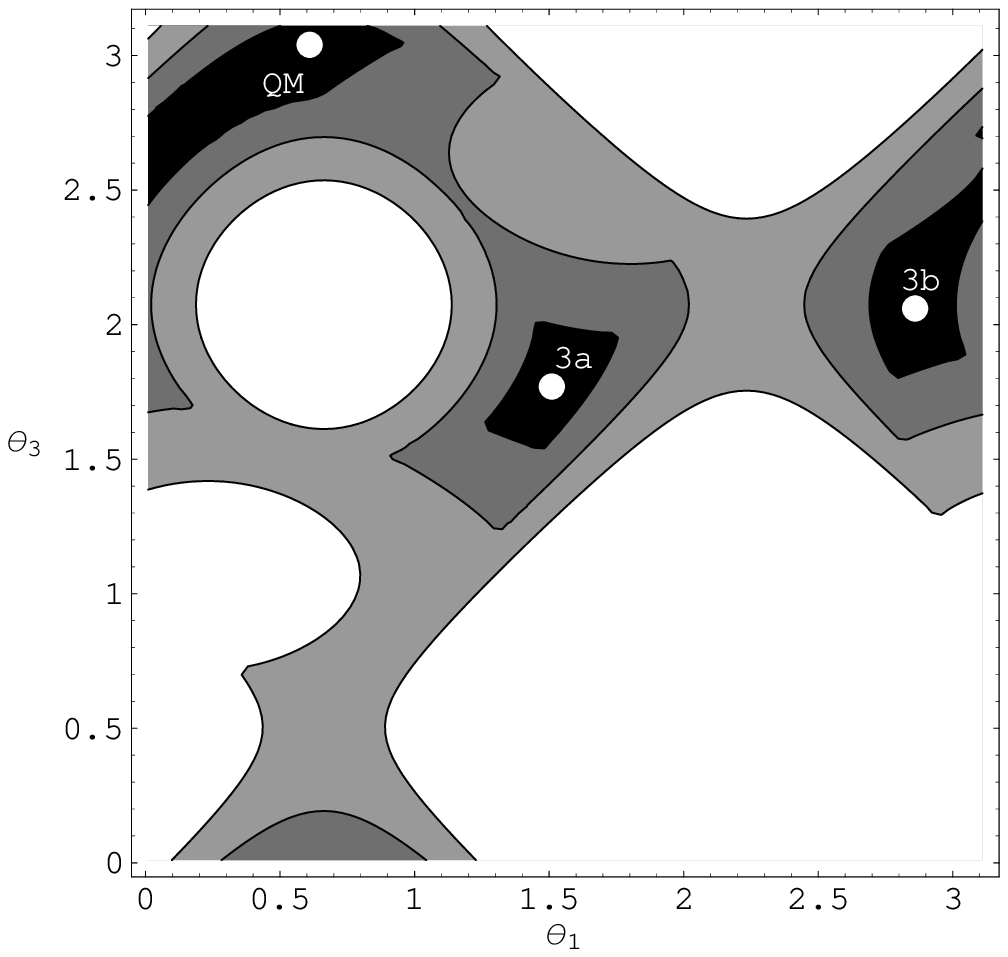}\hspace{1cm}
  \includegraphics[width=2in]{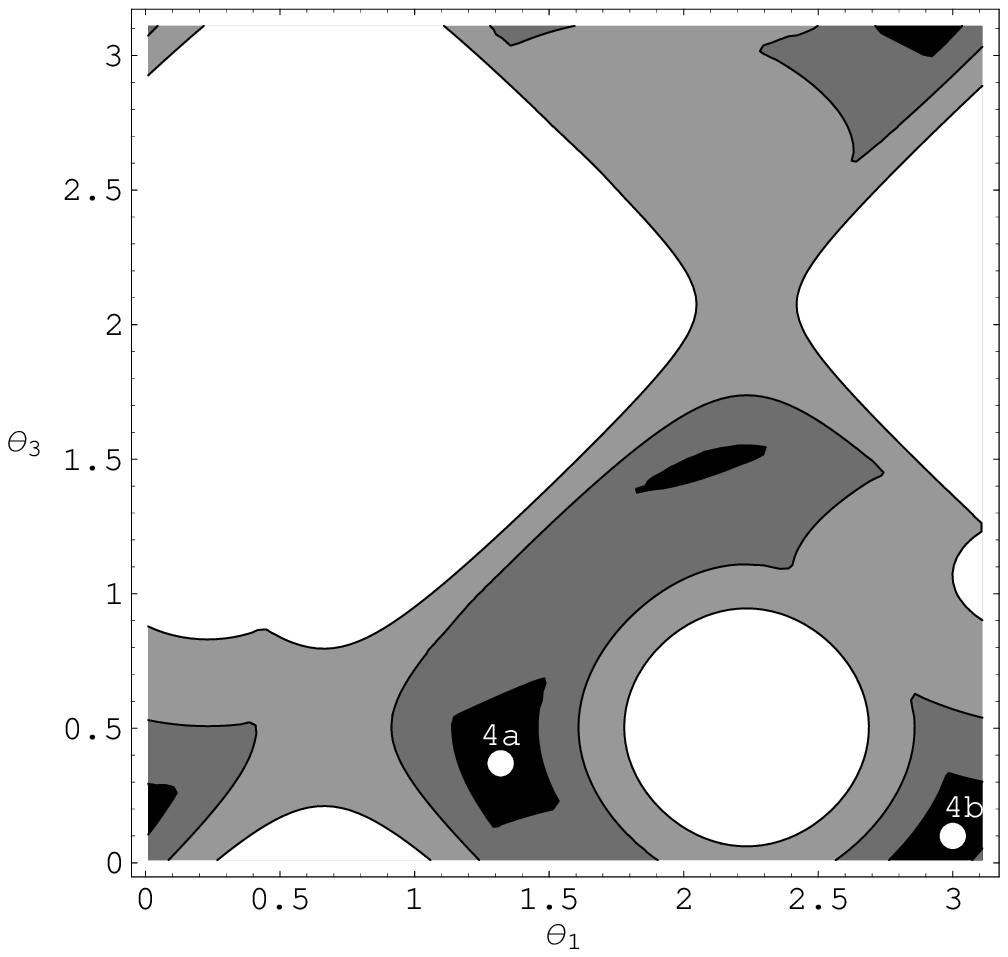}\\
\hspace{1cm} (c) \hspace{5cm} (d) \\
{\caption{
Regions in the $(\theta_{N1}, \theta_{N3})$ plane corresponding to 
$(C_2,C_3)$ close to the large $N_c$ values in Table 2, for each assignment:
(a) - \# 1, (b) - \# 2, (c) - \# 3, (d) - \# 4. The contour plots are given
by max$|C_{2,3}-C_{2,3}^{(0)}| = 50,100,150$ MeV.}}
\end{figure}

We present in Fig.~4 the regions allowed from such constraints, for each of the four
assignments introduced in Table 1. The regions in the $(\theta_{N1},\theta_{N3})$ plane 
are obtained from imposing the 
condition that the coefficients $C_{2,3}(\theta_1,\theta_3)$ do not differ by more
than 150 MeV from their large $N_c$ central values in Table 2. The
results for the assignment \#4 are given only for illustrative purposes, given that 
it is ruled out at leading order in $N_c$.
For example, the black areas in Fig.~4(a) denote the region in the $(\theta_{N1},
\theta_{N3})$ plane for which both $C_2$ and $C_3$ are within 50 MeV from their
central values in Table 2, corresponding to the assignment \#1. In the $C_{2,3}$ plane
(Fig.~2) this region is a square of sides 100 MeV centered on $(83 \; {\rm MeV},-188 \; {\rm MeV})$.
Each lighter shade of grey corresponds to an
additional 50 MeV. Similar contour plots are given in Figs.~4(b)-4(d) for each of 
the remaining three assignments in Table 1. Although the allowed regions appear
disjointed, they really form one continuous area because of the doubly periodic
conditions in the $(\theta_{N1},\theta_{N3})$ plane.

\begin{figure}
  \includegraphics[width=2.5in]{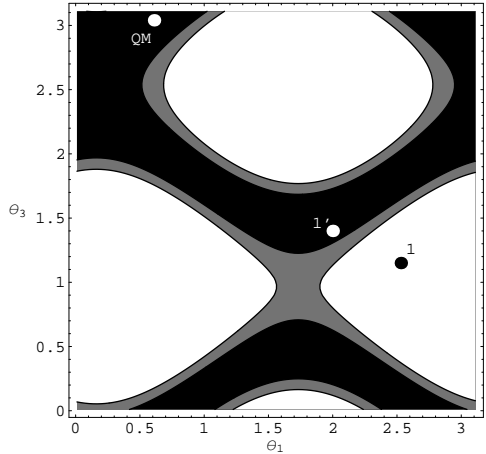} 
{\caption{
The constraints in the $(\theta_{N1}, \theta_{N3})$ plane following from
the $\Delta^*_{1/2}$ and $\Delta^*_{3/2}$ masses. The black (grey) regions
correspond to the $\Delta^*_{3/2}-\Delta^*_{1/2} = 75\pm 55$ ($75\pm 80$) MeV.}}
\end{figure}

It is not very illuminating to quote ranges of values for the coefficients
$C_k$ corresponding to the allowed regions in Figs.~4(a)-(d), since there
are significant correlations among coefficients. Instead we give in Table 4
their values at a few representative points in these plots, for each assignment.
These points correspond to minimal $1/N_c$ corrections to the coefficients
$C_{2,3}^{(0)}$ as determined in Table 2 at leading order in $N_c$. We quoted 
the coefficients $C_{1,6,7}$ in a form invariant under the ambiguity (\ref{ambig}). 
One can choose to vary $N_c C_1$ about its large $N_c$ values in Table 2 
over a range $\sim \Lambda$, which introduces a large uncertainty in the individual
values of $C_{1,6,7}$. On the other hand,
the combinations $N_c C_1^{\rm eff} = N_c C_1 - \frac12 C_7$ 
and $C_6^{\rm eff} = C_6 + C_7$ have smaller errors, of order $\Lambda/N_c$.

Finally, to compare with the results of \cite{CCGL} we consider also the
constraints from the $\Delta^*$ states. Although these states can not be included 
in a meaningful way in the large $N_c$ fit discussed at the beginning of this 
section, they can be used to constrain the $1/N_c$ determination.
Their masses are given explicitly by \cite{CCGL}
\bea
\Delta^*_{1/2} &=& N_c C_1 + \frac13 C_2 - \frac49 C_5 + \frac23 C_6 - \frac13 C_7\\
\Delta^*_{3/2} &=& N_c C_1 - \frac16 C_2 + \frac29 C_5 + \frac23 C_6 - \frac13 C_7\,.
\eea
Their splitting  $\Delta^*_{3/2} - \Delta^*_{1/2} = -\frac12 C_2 +\frac23 C_5 =75 \pm 80$
MeV gives an allowed region in the $(\theta_{N1},\theta_{N3})$ plane, 
shown in Fig.~5. This constraint includes the point $(\theta_{N1},\theta_{N3})_{QM} =
(0.61, 3.04)$ obtained from the quark model fit in \cite{CGKM}. While most of the
region corresponding to the assignment \#1 is excluded, this assignment is still
marginally allowed (see point $1'$ in Fig.~5).
Note that each point in the dark area of Fig.~5 gives an acceptable $O(1/N_c)$ 
mass matrix of the observed $N^*$, $\Delta^*$ states,
and generalizes the fit of \cite{CCGL}, which corresponds to the point QM and is
obtained by requiring agreement with the quark model mixing angles.

\bea\nonumber
\begin{array}{c|cc|rrr|rrrr}
\hline
 & \theta_{N1} & \theta_{N3} & N_cC_1^{\rm eff} & C_2 & C_3 & C_4 & C_5 & C_6^{\rm eff} & C_8 \\
\hline
\hline
\#1a & 2.18 & 1.18 & 1694 & 83 & -187 & -5 & 28 & -114 & -120  \\
\#1b & 2.93 & 0.50 & 1446 & 84 & -188 & -14 & -97 & 333 & -272  \\
\hline
\#2a & 0.19 & 0.58 & 1466 & 116 & -58 & -29 & -138 & 297 & -46 \\
\#2b & 2.72 & 1.34 & 1670 & 115 & -58 & -125 & -56 & -71 & 205 \\
\hline
\#3a & 1.51 & 1.77 & 1758  & -11 & 143 & 41 & 211 &  -230  & 360 \\
\#3b & 2.86 & 2.06  & 1611 & -13 & 142 &  -111  &  109  & 35  & 451  \\
\mbox{QM} & 0.61 & 3.04 & 1410 & -35 & 123 & 85 & 84 & 398 & 49 \\
\hline
\#4a & 1.32 & 0.37 & 1508  & 2 & -110 & 186 & 97 &  222  & -360 \\
\#4b & 3.00 & 0.10  & 1379 & 2 & -110 &  22  &  0  & 454  & -245  \\
\hline 
\end{array}
\eea
\begin{quote}
Table 4. Coefficients (in MeV) and angles (in rad) for the points shown in 
Fig.~4(a)-(d). We also show the quark model point (QM) corresponding to 
the angles $\theta_{N1,N3}$ used as inputs in \cite{CCGL}.
\end{quote}

\section{Conclusions}

We studied in this paper the mass spectrum of the $L=1$ orbitally excited 
nonstrange baryons in the $1/N_c$ expansion. In the quark model with spin-flavor
independent quark forces, these states fall into the {\bf 70} representation of 
SU(6) \cite{su6}. While large $N_c$ QCD arguments can be used to reproduce many of
the predictions of the quark model for these states in a model-independent
way \cite{PY1}, they also predict
a nontrivial structure of the mass spectrum which  breaks spin-flavor 
symmetry \cite{JG}. Still, a significant amount of symmetry remains, in the form 
of the contracted $SU(2N_f)$ symmetry of large $N_c$ QCD \cite{DJM1,LuMa,DJM2}. 
This symmetry
predicts that in the large $N_c$ limit, the $L=1$ orbitally excited baryons
arrange themselves into 3 infinite towers of degenerate states \cite{PY1,PY2}. 
These towers are labeled by one integer $T=0,1,2$, and contain all states 
which satisfy $|J-I| \leq T$.

In a parallel development, the structure of the mass spectrum of the excited baryons 
was studied in the
$1/N_c$ expansion \cite{JG,CCGL1,CCGL,SGS,su3prd} using the different but equivalent 
approach of quark operators. In this approach, the mass operator is written as the
most general sum of flavor singlet Lorentz scalar operators which contribute to a given
order in $1/N_c$. Using such method, the mass spectrum of the $L=1$ baryons was studied
in both $SU(2)$ and $SU(3)$ flavor, up to subleading order $O(1/N^2_c)$.

In this paper we show explicitly how the predictions of the $SU(4)_c$ symmetry
for nonstrange excited baryons follow from the quark mass operator approach in the
large $N_c$ limit. Apart from recovering the known structure of the mass spectrum 
in the symmetric limit, this approach allows also the study of the $1/N_c$ symmetry 
breaking corrections.
We discuss the special status of the excited $\Delta^*$ states in the large $N_c$ expansion;
for $N_c \geq 3$ there are 8 $\Delta^*$ states (vs. only 2 at $N_c=3$), and their 
mixing has to be taken into account in order to reproduce the large $N_c$ symmetry predictions.
This casts some doubt on the ability of the $1/N_c$ expansion to correctly describe
their properties.

The main aim of our paper is to answer the two (related) questions: 
a) are the predictions of large 
$N_c$ $SU(2N_f)_c$ symmetry still visible in the observed mass spectrum of the excited
baryons? and b) to determine the values of the coefficients of the various operators in
the mass operator. 

At the first sight, the structure of the observed mass spectrum of the $N^*$ states 
is very similar to the expected set of towers, with states almost degenerate in pairs
$\{N_{1/2}(1535), N_{3/2}(1520)\}$, and $\{N_{3/2}(1700), N_{5/2}(1675)\}$. This 
corresponds to a special assignment of the physical states into $SU(4)_c$ representations
(\#1), and 
is supported by the full $1/N_c$ analysis. This assignment allows for small coefficients 
for the $O(1/N_c)$ operators $C_{4-8}$ in the mass matrix (see Table 4, \#1a). 
Even after including the excited $\Delta^*$ (see Fig.~5), it is marginally still allowed,
in a region around $(\theta_{N1},\theta_{N3})_{\rm 1'} \sim (2.0, 1.4)$.
We conclude that the assignment \#1 is still allowed within the present experimental 
uncertainties.

Allowing for larger $1/N_c$ corrections, 3 other assignments of the observed states into
large $N_c$ towers are possible (see Table 1). We show that model-independent constraints 
in the large $N_c$ limit rule out one of them (\#4). The remaining two 
assignments require (as expected) relatively large coefficients of the $O(N_c^{-1})$ 
operators (see Table 4), depending on the precise
position in the $(\theta_{N1}, \theta_{N3})$ parameter space. Beyond leading order in $N_c$ 
there is a double continuum of coefficients $C_k(\theta_{N1}, \theta_{N3})$ 
(with one additional ambiguity when 
excluding the $\Delta^*$ states).  We use dimensional analysis to study the 
regions in the parameter space where the $1/N_c$ expansion is well behaved (see Figs.~4).

We find that, despite requiring large $O(N_c^{-1})$ coefficients, the assignment \#3 is 
currently favored by the constraints from the $\Delta^*$ mass splittings.
This agrees with the results of the fit in \cite{CCGL}, where a special solution for the
coefficients of the mass operator was determined using as inputs the masses and mixing angles 
of the orbitally excited $N^*$, $\Delta^*$ states. 
The specific pattern of the operator coefficients in this fit 
was invoked in the literature \cite{OPE,Riska} as supporting a model of interquark forces 
based on Goldstone boson exchange (as opposed to one-gluon exchange). Although clearly 
interesting, we find that it is premature to draw any such conclusions based on the 
large $N_c$ expansion. As noted above, even in the large $N_c$ limit, the mass operator 
can be determined only with discrete ambiguities; at subleading order this becomes a 
double (triple) continuous ambiguity, depending on whether the $\Delta^*$ states are (are not) 
included. 

Further progress towards resolving this multiple ambiguity can be made by considering also the
strong couplings of these states. In the large $N_c$ limit, these were shown to be related to the 
assignment of the physical states into large $N_c$ towers \cite{PY1}. Unfortunately, the precision
of the present data is still not sufficient to resolve the ambiguity \cite{PY2}. Progress can also
come from lattice QCD, from which first results on excited nucleons are already available \cite{lattice}.
As discussed in Sec.~III, the specific assignment realized in Nature depends on the values of the
large $N_c$ coefficients $C_{2,3}^{(0)}$, which could be computed in a lattice simulation. Finally,
improvements in the experimental uncertainties in the masses of these states (which will be available
from Jefferson Lab in the next future) could sharpen the
constraints presented here. Clearly, more than forty years after their discovery,  the orbitally excited 
baryons remain an active and exciting field of study.

\vspace{0.3cm}
{\em Note added:} After completing this paper we became aware of similar work on the leading $N_c$
mass spectrum of excited baryons done in \cite{CoLe}.

\vspace{0.5cm}
We are grateful to Jos\'e Goity for useful discussions, to Elizabeth Jenkins for
comments on the manuscript, and to the organizers of the
workshop `The Phenomenology of Large $N_c$ QCD', Tempe, AZ 2002, where this work
was initiated. D.P. acknowledges the hospitality of the Jefferson Laboratory and the
Duke University physics department during the final phase of this
work. This work has been supported by the DOE and NSF under Grants No. DOE-FG03-97ER40546 (D.P.),
NSF PHY-9733343, DOE DE-AC05-84ER40150 and DOE-FG02-96ER40945 (C.S.).

\appendix
\section{Coefficients}

We give here the explicit results for the coefficients 
$C_{1-8}(\theta_{N1},\theta_{N3})$ used in Sec.~III. We denote in these expressions 
$t_1 = \tan \theta_{N1}$ and $t_3 = \tan \theta_{N3}$.
\bea
C_2(\theta_{N1},\theta_{N3}) &=& \frac{1}{3(1+t_1^2)(1+t_3^2)}
\left\{ (1+t_3^2)(-2t_1^2+t_1) N(1650)
+ (1+t_1^2)(\sqrt{10}t_3 + 2t_3^2)  N(1700)
\right.\nonumber\\
& &\left. - (1+t_1^2)(-2+\sqrt{10} t_3) N(1520)
- (2+t_1)(1+t_3^2) N(1535)\right\}\\
C_3(\theta_{N1},\theta_{N3}) &=& \frac{2}{15(1+t_1^2)(1+t_3^2)}
\left\{ 5(1+4t_1)(1+t_3^2)  N(1650)\right.\nonumber\\
& &\left. + ( -8 - 4\sqrt{10} t_3 ) (1+t_1^2) N(1700)
 + (4\sqrt{10} t_3 - 8 t_3^2) (1+t_1^2)N(1520)\right.\nonumber\\
& &\left. + (-20 t_1 + 5 t_1^2)(1+ t_3^2 ) N(1535)
+ 3(1+ t_1^2)(1+t_3^2) N(1675)
\right\}\\
C_4(\theta_{N1},\theta_{N3}) &=& -\frac{1}{15(1+t_1^2)(1+t_3^2)}
\left\{ 5(1-4t_1^2)(1+t_3^2)  N(1650)\right.\nonumber\\
& &\left. -9 (1+t_1^2)(1+t_3^2) N(1675)
 + 4(1+t_1^2)(1+5t_3^2) N(1700)\right.\nonumber\\
& &\left. + 5(-4 + t_1^2)(1+ t_3^2 ) N(1535)
+ 4(1+ t_1^2)(5+t_3^2) N(1520)
\right\}\\
C_5(\theta_{N1},\theta_{N3}) &=& -\frac{1}{20(1+t_1^2)(1+t_3^2)}
\left\{ 5(3+2t_1-4t_1^2)(1+t_3^2)  N(1650)\right.\nonumber\\
& &\left. -27 (1+t_1^2)(1+t_3^2) N(1675)
 + (1+t_1^2)(12 + 10\sqrt{10}t_3 + 20t_3^2) N(1700)\right.\\
& &\left. + 2(10 - 5\sqrt{10} t_3 + 6t_3^2)(1+ t_1^2 ) N(1520)
+ 5(1+ t_3^2)(-4-2t_1+3t_1^2) N(1535)
\right\}\nonumber\\
C_6(\theta_{N1},\theta_{N3}) &+& 6 C_1(\theta_{N1},\theta_{N3}) =
 \frac{1}{6(1+t_1^2)(1+t_3^2)}
\left\{ (2+t_1^2)(1+t_3^2)  N(1535)\right.\nonumber\\
& &\left. + (1+2t_1^2)(1+t_3^2) N(1650)
 + 3(1+t_1^2)(1 + t_3^2) N(1675)\right.\nonumber\\
& &\left. + 2(1 + 2t_3^2)(1+ t_1^2 ) N(1700)
+ 2(2+ t_3^2)(1+t_1^2) N(1520)
\right\}\\
C_7(\theta_{N1},\theta_{N3}) &-& 6 C_1(\theta_{N1},\theta_{N3}) =
 \frac{1}{3(1+t_1^2)(1+t_3^2)}
\left\{ (-4+t_1^2)(1+t_3^2)  N(1535)\right.\nonumber\\
& &\left. + (1-4t_1^2)(1+t_3^2) N(1650)
 + 3(1+t_1^2)(1 + t_3^2) N(1675)\right.\nonumber\\
& &\left. + 2(1 - 4t_3^2)(1+ t_1^2 ) N(1700)
+ 2(-4 + t_3^2)(1+t_1^2) N(1520)
\right\}\\
C_8(\theta_{N1},\theta_{N3}) &=& -\frac{1}{10(1+t_1^2)(1+t_3^2)}
\left\{ (40 + 8\sqrt{10} t_3)(1+t_1^2)  N(1700)\right.\nonumber\\
& &\left. + (40t_1 - 25t_1^2)(1+t_3^2) N(1535)
 + (1+t_1^2)(-8\sqrt{10} t_3 + 40 t_3^2) N(1520)\right.\nonumber\\
& &\left. - 5(5 + 8t_1)(1+ t_3^2 ) N(1650)
- 15(1+ t_1^2)(1+t_3^2) N(1675)
\right\}\,.
\eea

\end{document}